\documentclass[prb,showpacs,twocolumn,aps,floatfix]{revtex4}
\usepackage{graphicx}
\begin{document}

\title{Elastic anomalies in {\sl HoNi}$_{2}${\sl B}$_{2}${\sl C} single crystals}
\author{V.D. Fil$^{1}$, A. Knigavko$^{2}$, A.N. Zholobenko$^{1}$, E.-M. Choi$^{3}$,
and S.-I. Lee$^{3}$.}
\affiliation
{$^{1}$B.Verkin Institute for Low Temperature Physics and Engineering,
National Academy of Sciences of Ukraine, pr. Lenina 47, Kharkov 61103,
Ukraine\\
$^{2}$Department of Physics and Astronomy, McMaster University, Hamilton,
ON, L8S 4M1, Canada\\
$^{3}$National Creative Research Initiative Center for Superconductivity and
Department of Physics, Pohang University of Science and Technology, Pohang
790-784, Republic of Korea}

\begin{abstract}
We have measured temperature and magnetic field dependencies of the sound
velocities and the sound attenuation in {\sl HoNi}$_{2}${\sl B}$_{2}${\sl C}
single crystals. The main result is a huge softening the velocity of $C_{66}$
mode due to a cooperative Jahn-Teller effect, resulting in a
tetragonal-orthorhombic structural phase transition. Anomalies in the
behavior of the $C_{66}$ mode through various magnetic phase transitions
permit us to revise the low temperature H--T phase diagrams of this compound.
\end{abstract}
\maketitle


Holmium borocarbide {\sl HoNi}$_{2}${\sl B}$_{2}${\sl C} is one of the most
interesting members of the rare earth borocarbide family. The list of
phenomena that have been found in this compound at low temperatures contains
many of the mainstream research issues of solid state physics. Among them are
the coexistence of superconductivity and magnetism, occurrence of multiple
magnetic phases, including incommensurate spiral structures, and reentrant
superconductivity (see \cite{muller} for a review). In this
paper we uncover one more facet of complexity of {\sl HoNi}$_{2}${\sl B}$%
_{2} ${\sl C}: the presence of strong Jahn--Teller interactions.

Over the past decade almost all of the available arsenal of experimental tools has
been used in the studies of borocarbides. However it seems that
ultrasound methods did not receive the proper attention, and as a result the
acoustic properties of these compounds are mostly ``terra incognito'' at
present. In this work we have performed acoustic investigation of {\sl HoNi}$%
_{2}${\sl B}$_{2}${\sl C} single crystals and obtained valuable data that
are complementary t o the known results. In particular, we have discovered
a considerable softening of the $C_{66}$ elastic modulus that starts at
temperatures as high as 100K. This is indicative of the strong Jahn--Teller
interaction present in this compound. This interaction is the driving force
of the tetragonal--orthorhombic structural phase transition \cite{dist-neurt}.
The magnetic phase transitions are accompanied by elastic anomalies which are 
most pronounced for the $C_{66}$ mode. By analyzing these features we were able
to construct the H-T phase diagrams that added a few details to the existing
phase diagrams (see for example \cite{specific-heat}).

\begin{figure}[th]
\includegraphics[width=8.5cm,angle=0]{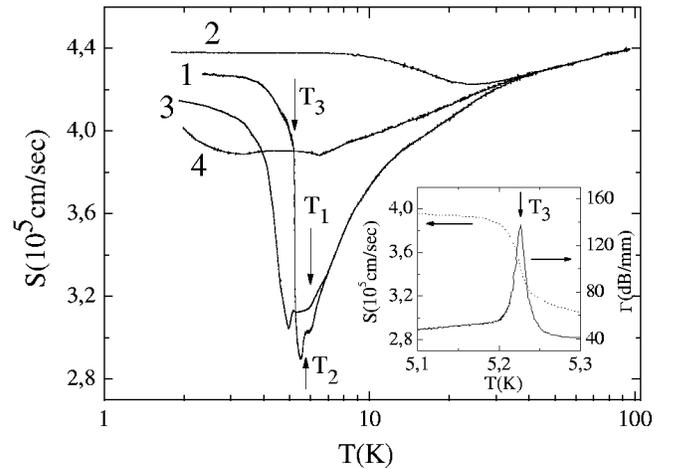}
\caption{
Temperature dependence of the velocity of the $C_{66}$ mode. 
Curve 1: $H=0$, curves 2, 3 and 4:  $H=4T$ oriented along $[110]$, $[001]$ 
and $[100]$ correspondingly. Arrows indicate the anomalies for $H=0$.  Insert:
the attenuation (full line, right axis) and velocity (dotted line, left axis) of 
the $C_{66}$ mode near $5.2K$ for $H=0.$
}
\end{figure}

Single crystals of {\sl HoNi}$_{2}${\sl B}$_{2}${\sl C} were produced using
the method described previously \cite{crystal-grows}. Samples had the shape of
plates of 0.2 mm thickness. The c axis of the crystal was oriented
perpendicular to the surface of the plate. The technique of working with
submillimeter samples and the method of phase measurements are presented 
elsewhere \cite{ulta-sound-exp}. The characteristic feature of the method 
is the utilization of an electronically controlled phase shifter with a dynamical 
range that is practically unlimited. This, along with the use of a digital phase 
measuring device, allows to keep a very high resolution for large changes in the 
sound velocity. 
While the behavior of all pure modes of the crystal have been studied, we devote
this paper mainly to the analysis of the most informative $C_{66}$ mode. The 
anomalies of the sound velocity of the $C_{44}$ mode are also briefly discussed
at the end. 

Shown in Fig. 1 are the temperature dependences of the sound velocity, $S$,
of the $C_{66}$ mode both in the absence of an external magnetic field $H$
and when the field is oriented along main crystal directions. For $H=0$
(curve 1) $S$ exhibits a considerable, about 35\%, softening that starts
showing up long before the onset of magnetic ordering found at 
$T\sim10K$ by neutron diffraction experiments \cite{dist-neurt,neut-mag-order}. 
At $T_{cr}=5.23K$ there is a return to a stiffer state marked by a jump in $S$. 
If a magnetic field of $H=4T$ is applied in the basal plane (curves 2
and 4) the magnitude of the softening is strongly reduced, though not
eliminated completely. On the other hand, $S$ is affected very little if the
field is applied along the $c$ axis (curve 3).

\begin{figure}[th]
\includegraphics[width=8.5cm,angle=0]{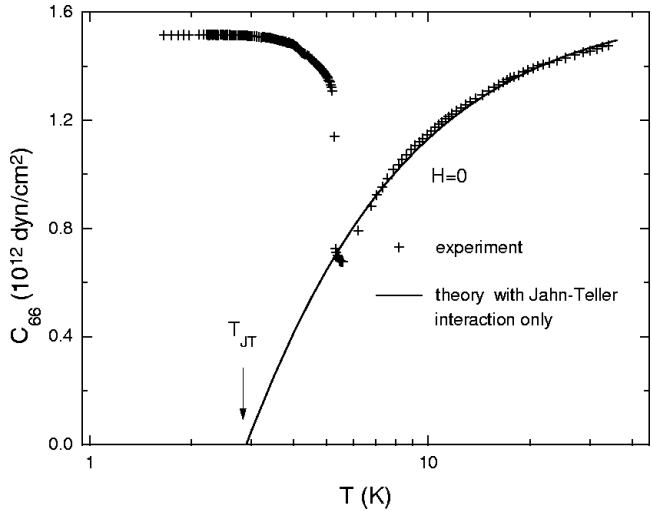}
\caption{
Temperature dependence of the $C_{66}$ elastic modulus for $H=0$. Crosses are
the experimental points, solid line is the theoretical result.
}
\end{figure}

A wide temperature interval in which a softening of elastic constants takes
place is the fingerprint of the Jahn--Teller interaction \cite{luthi1}. The
elastic constant of the lattice itself (background elastic constant $C^{(0)}$)
increases very slowly with decreasing temperature. Much stronger temperature
dependence comes from partially filled inner electronic shells of rare earth
ions. For {\sl HoNi}$_{2}${\sl B}$_{2}${\sl C} the crystal field splitting
of the lowest J--multiplet of $Ho^{3+}$ ion is about 200K \cite
{cristal-field}, which explains why we do not see the pure background
elasticity even at the highest temperatures used in our experiments (about
100K). The localized electronic degrees of freedom are coupled to an
extended degrees of freedom associated with the lattice. The elastic modulus
of interest can be calculated as follows (see, for example \cite{luthi1}):
\begin{eqnarray}
C_{66}=C_{66}^{(0)}
-{T\over v_{cell}}{d^2 \over d \varepsilon_6^2}
\left[
\log\sum_i \exp\left({-E_i(\varepsilon_6)\over T} \right)
\right]_{\varepsilon_6=0}
\nonumber
\end{eqnarray}
where $v_{cell}$ is the unit cell volume and 
$E_i$ are the eigenvalues of the single ion Hamiltonian $H_{ion}=V_{CF}+V_{JT}$
which operates within the lowest multiplet of $Ho^{3+}$ ion with J=8. 
The the crystal field Hamiltonian $V_{CF}$ has the form standard 
for tetragonal symmetry with the parameters determined by Gasser {\it et al.} 
\cite{cristal-field} The magnetic-ion--lattice (Jahn--Teller) interaction is taken 
in the simplest form $V_{JT}=-g \varepsilon _{6}O_{2}^{-2}$, 
where $\varepsilon _{6}\equiv \varepsilon _{xy}$ is the relevant component of the 
strain tensor and $O_{2}^{-2}$ is the lowest order electronic quadrupole operator 
to which this strain couples by symmetry. 
Both $C_{66}^{(0)}$, assumed temperature independent, and the unknown coupling constant $g$ were determined from a fitting procedure in the 
temperature range where the system is paramagnetic. 
In Fig. 2 we show  and the best fit (solid line), which corresponds to $g=2.24$ meV/ion and $C_{66}^{(0)}=1.63 \times 10^{12}$dyn/cm$^2$, along with the experimental points (crosses) for $H=0$. The agreement is quite good. The
steep increase of $C_{66}(T)$ at $T_{cr}=5.23K$ is due to interactions with
magnetic degrees of freedom, 
which we do not attempt to account for in this work.

The next conclusion is that Jahn--Teller interaction is in fact quite
strong. If there were no other interactions in the system the crystal
lattice would become unstable at about $T_{JT}=2.89K$ as indicated by
the vanishing of the calculated $C_{66}$ (see Fig. 2). Therefore, the
Jahn--Teller interaction is the driving force of the structural phase
transition from the tetragonal to orthorhombic phase that most probably takes
place simultaneously with the stabilization of collinear antiferromagnetic
ordering at $T_{cr}=5.23K$. Orthorhombic distortions of the crystal lattice
have been observed previously at about 2K \cite{dist-neurt} and were
associated with the magnetostriction.

\begin{figure}[th]
\includegraphics[width=8.5cm,angle=0]{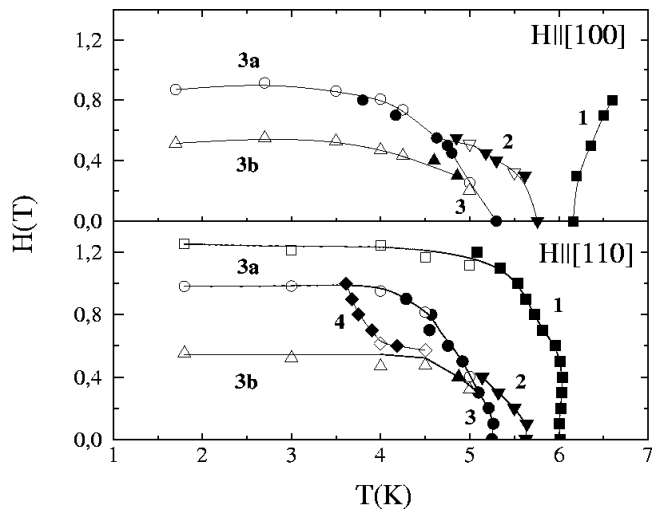}
\caption{
Phase $H-T$ diagrams for $H||[100]$ (top panel) and $H||[110]$ (bottom panel). 
Points belonging to different branches are denoted by symbols of 
different shape; lines are guidance for eyes. 
Full symbols -- results of $T$--sweeps with fixed $H$,
open symbols -- results of $H$--sweeps with fixed $T$.
}
\end{figure}

Neither the sound velocity nor the sound attenuation, $\Gamma $, develop
any visible anomalies at the superconducting phase transition ($T_{c}\simeq
8K$ \cite{muller}) in our experiments. At the same time magnetic phase
transformations can be detected quite easily with the behavior of the $C_{66}$
mode. The typical shapes of the anomalies in the sound velocity and the
sound attenuation are clearly seen in Fig. 1. The characteristic features
are different for different phase transformations. For the sound velocity
there is a change in the slope (jump in the derivative) at temperatures $%
T_{1}$, while at $T_{2}$ there is a jump-like decrease and at $T_{3}$ there
is a jump-like increase. 
Usually, these kinds of anomalies in the sound velocity are
accompanied by an increase of the sound attenuation, which can be quite large
in some cases (for example at $T_{3}$, see the insert in Fig. 1). In
principle, by the type of anomaly one can determine the nature of the
corresponding phase transformation. For instance, the change in slope of the
sound velocity at $T_{1}$ corresponds to a second order phase transition of
the improper segnetoelastic type. 

Leaving such an analysis for a future publication we present in Fig. 3 
the obtained $H-T$ phase diagrams for $H||[100]$ (top panel) 
and $H||[110]$ (bottom panel).
Note that we used two types of measurements: temperature sweeps with fixed 
magnetic field  and field sweeps with fixed temperature. In all cases the 
external parameter was kept increasing during sets of sweeps.

\begin{figure}[th]
\includegraphics[width=8.5cm,angle=0]{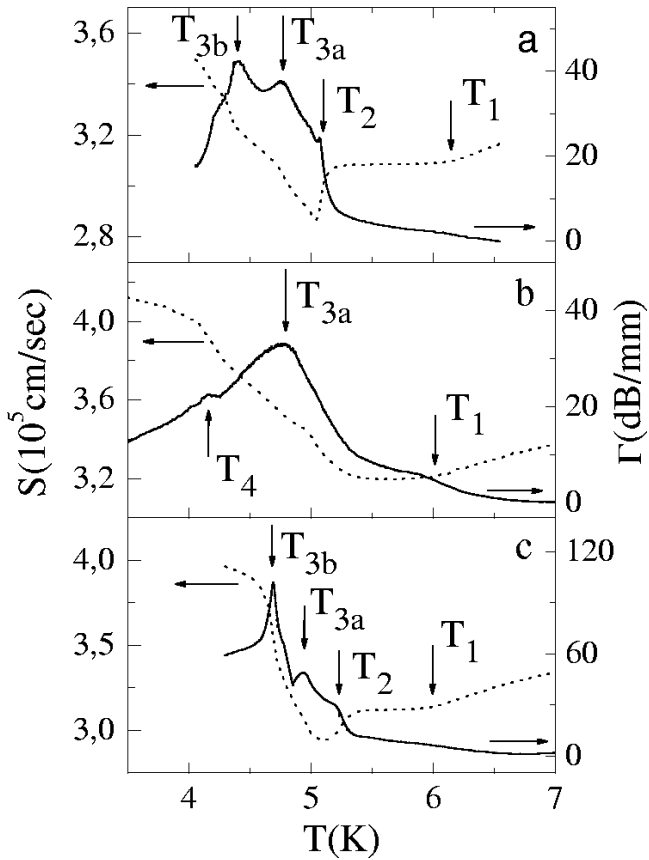}
\caption{
Temperature dependence of the attenuation (full lines, right axis) and 
velocity (dotted lines, left axis) of the $C_{66}$ mode for 
a) $H(0.45T)||[100]$, b) $H(0.6T)||[110]$ and c) $H(3T)||[001]$.
}
\end{figure}

Let us discuss the case of $H||[100]$ (see Fig. 3, top panel) in more detail.
At $H=0$ the temperatures of the magnetic phase transitions practically coincide
with those found in the literature \cite{muller}. The location of line 1
is close to that of the other published data as well. Line 3
represents the position of a jump-like increase of the sound velocity, which
is steep at $H=0$ but smears considerably as $H$ grows. The attenuation
always goes through a maximum on line 3 (see insert on Fig. 1). At $%
H\simeq 0.4T$ line 3 splits forming a critical point. Shown in Fig. 4a
are the $S(T)$ and $\Gamma (T)$ curves at $H=0.45T$ where the splitting of
the line 3 is obvious. In the literature lines 3 and 3b are associated
with the boundary of a commensurate antiferromagnetic phase. Both line 3a
and line 3b represent an increase of the sound velocity in the form of a
smeared step. They are possibly better detected as peaks in the attenuation
(see Fig. 4a, right vertical axis). Although it is in principle possible
that lines 3b and 3b stay separate from each other as the magnetic field
decreases all the way down to zero, we were not able to observe any
indications of such behavior. Line 3 is not split as $H\rightarrow 0$
within the accuracy of our experiments (see insert in Fig. 1).

\begin{figure}[ht]
\includegraphics[width=8.5cm,angle=0]{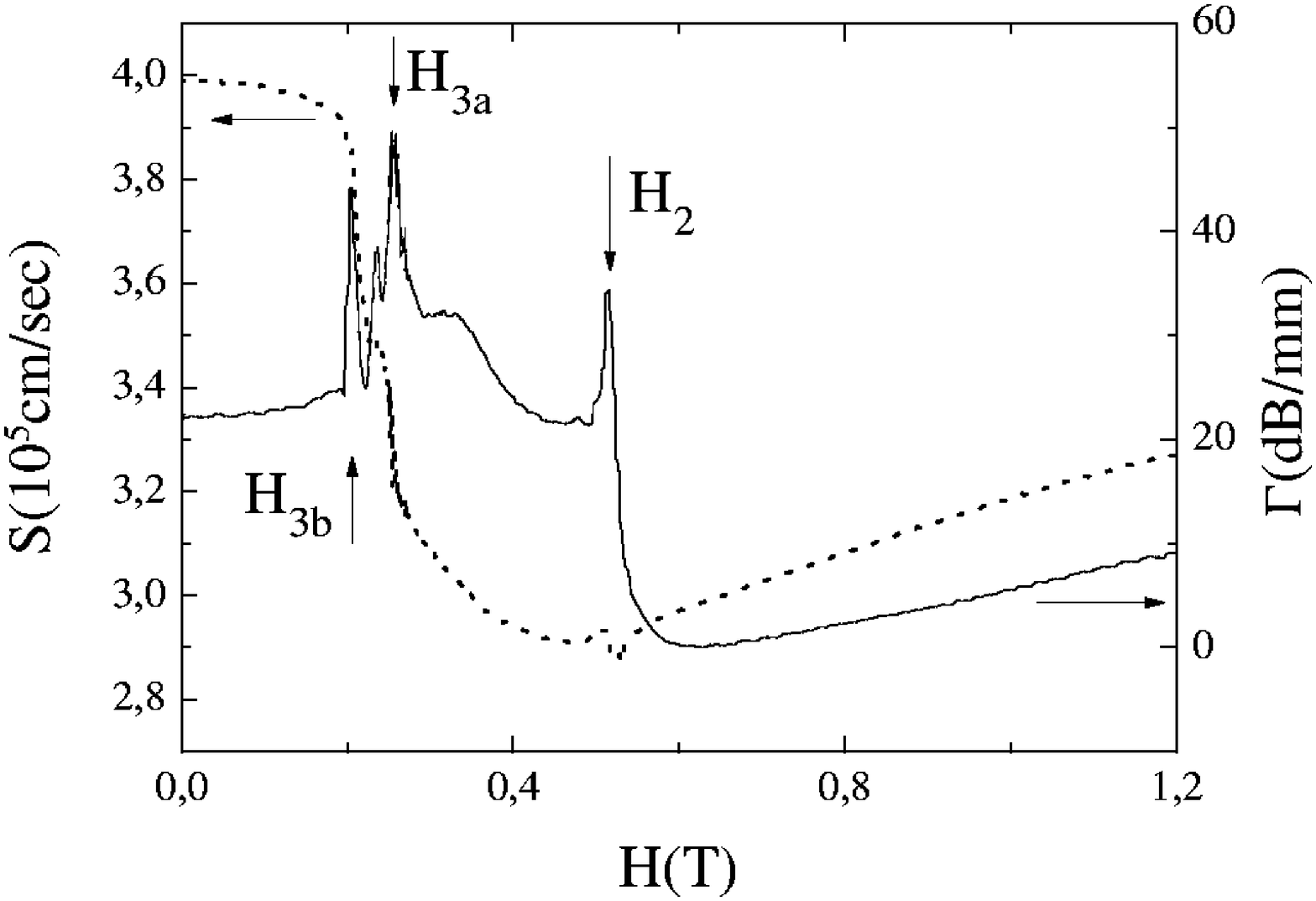}
\caption{
Magnetic field dependence of the attenuation (full lines, right axis) and
velocity (dotted lines, left axis) of the $C_{66}$ mode at $T=5K$ for $H||[100]$.
}
\end{figure}

The anomaly represented by line 2 is a step-like decrease of the sound
velocity. At $H||[100]$ line 2 approaches line 3a as field increases, 
but disappears further on. Over some range of magnetic fields 
line 2 coexists with lines 3a and 3b (see Fig. 3, top panel). This is
demonstrated in Fig 4a where all three peaks in the attenuation $\Gamma $
are clearly distinguishable. The simultaneous presence of all these three
anomalies can also be observed on an appropriate vertical cut of the $H-T$
plane in the top panel of Fig. 3. The magnetic field dependence of $S$ and 
$\Gamma $ at $T=5K,$ shown in Fig 5, confirms the coexistence of lines 2, 
3a and 3b unambiguously.

In general, similar systematics holds true for $H||[110]$ as well (see
Fig. 3, bottom panel). However, in the fields larger than the value at which
lines 2 and 3a merge one can see a new feature on $S(T)$ and $\Gamma (T)$
curves (look up $T_{4}$ in Fig. 4b). This suggests the existence of one more
line, line 4 in the bottom panel of Fig. 3, which could in fact be a
continuation of line 2.

For $H||[001]$ the $H-T$ phase diagram essentially coincides with the one
published in the literature \cite{muller}. We do not present it again in
this paper, but comment that in high fields line 3 is possibly split as
well. Indeed, in fields as high as 3T (see Fig. 4c), we can readily define 
$T_{3a}$ and $T_{3b}$ on the attenuation temperature dependence curve, 
$\Gamma (T)$.

\begin{figure}[th]
\includegraphics[width=8.5cm,angle=0]{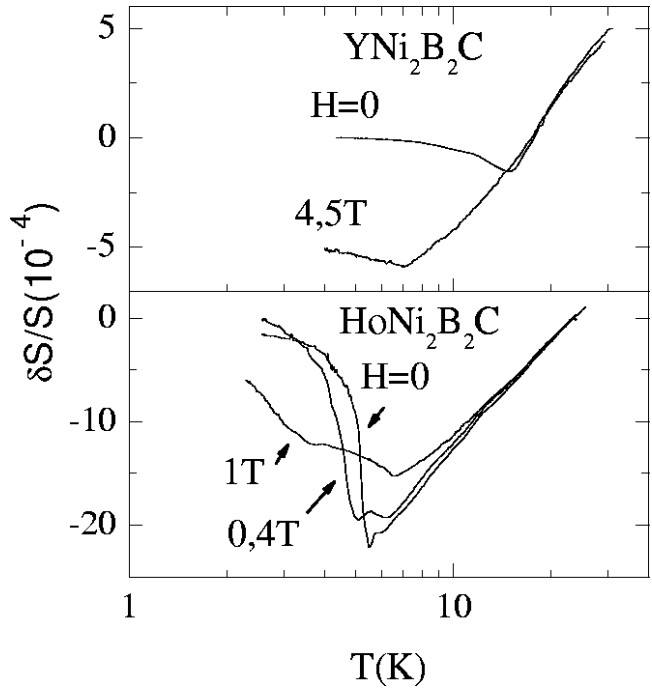}
\caption{
Comparison of the temperature dependence of the $C_{44}$ mode
velocity for {\sl YNi}$_{2}${\sl B}$_{2}${\sl C} (top panel; $q||[001],H||[001]$) 
and {\sl HoNi}$_{2}${\sl B}$_{2}${\sl C} (bottom panel; $q||[001],H||[100]$).
}
\end{figure}

Fig. 6 shows the temperature dependences of the velocity of the $C_{44}$ mode 
in both {\sl Ho} and {\sl Y} borocarbides that also exhibits slight softening.
Notice that the magnitude of the effect for this mode is three orders of
magnitude smaller than for the $C_{66}$ mode in the {\sl Ho} compound (see Fig.
1). The $Y^{3+}$ ions in the other compound have all electronic shells
completely filled, and therefore do not couple to the lattice by means of
the mechanism discussed above. A softening of the $C_{44}$ mode velocity in 
{\sl YNi}$_{2}${\sl B}$_{2}${\sl C} for $H=0$ has been observed recently by
Isida {\it et al} \cite{isida} who associated it with the presence of a
nesting feature of the Fermi surface. 

Such anomalies have attracted a lot of attention in the 1970s in connection
with superconductors of the A-15 crystal structure as a possible source of low
temperature structural transformations \cite{testardi}. The theory \cite
{gorkov} predicts a logarithmic dependence of the active elastic modulus on 
temperature (for a given geometry). This is satisfied reasonably well
for both compounds (see Fig. 6), providing the evidence of a measurable
coupling of itinerant electrons to the lattice. In borocarbides this
coupling is much weaker than in A-15 compounds, which could be due to a
smaller contribution of 1D bands to the electronic spectrum. 

Apart from a similar temperature dependence in the
range of temperatures where softening takes place, the behavior of the $C_{44}$ 
elastic modulus in {\sl Y} and {\sl Ho} borocarbides with further
decrease of temperature is very different. In the former compound the
lattice softening stops upon stabilization of the superconducting phase,
including the mixed state in magnetic fields (Fig. 6, top panel). In the
latter compound the lattice starts growing stiffer only after magnetic
ordering occurs while the onset of superconductivity does not produce any
visible effect.

In summary, we demonstrated the presence of a Jahn--Teller interaction in 
{\sl HoNi}$_{2}${\sl B}$_{2}${\sl C} which is sufficiently strong to make
the high temperature tetragonal crystal structure unstable (a cooperative
Jahn--Teller effect). 
The details of $H-T$ phase diagram were clarified. We presented
evidence in favor of the existence of critical points where several
phase transition lines merge or perhaps intersect.

This study was supported in part by CRDF Foundation (Grant UP1-2566-KN-03).
The work of AK was supported by the Natural Science and Engineering Research
Council of Canada (NSERC) and the Canadian Institute for Advanced Research
(CIAR).


\end{document}